# ACOUSTIC CYBERSECURITY: EXPLOITING VOICE-ACTIVATED SYSTEMS

Forrest McKee[1] and David Noever[2]

PeopleTec, 4901-D Corporate Drive, Huntsville, AL, USA, 35805

[1]forrest.mckee@peopletec.com  [2] david.noever@peopletec.com

## ABSTRACT

*In this study, we investigate the emerging threat of inaudible acoustic attacks targeting digital voice assistants, a critical concern given their projected prevalence to exceed the global population by 2024. Our research extends the feasibility of these attacks across various platforms like Amazon's Alexa, Android, iOS, and Cortana, revealing significant vulnerabilities in smart devices. The twelve attack vectors identified include successful manipulation of smart home devices and automotive systems, potential breaches in military communication, and challenges in critical infrastructure security. We quantitatively show that attack success rates hover around 60%, with the ability to activate devices remotely from over 100 feet away. Additionally, these attacks threaten critical infrastructure, emphasizing the need for multifaceted defensive strategies combining acoustic shielding, advanced signal processing, machine learning, and robust user authentication to mitigate these risks.*

## 1. INTRODUCTION

By 2024, industry forecasts suggest that the number of digital voice assistants will reach 8.4 billion units – exceeding the world's population [1]. The advent of voice-activated systems such as Amazon's Alexa ecosystem, with its myriad of software and hardware combinations (over 50 reported), offers users convenience while presenting an attractive target for malicious actors [2-3]. With over 300 million smart home devices connected to Alexa, including lighting, thermostats, cameras, robotic vacuums, and locks, the potential for compromise is substantial [4]. The Android and iOS platforms collectively encompass a staggering number of models and applications, with Android alone boasting 24,000 device models and over a million apps on the Google Play Store [5]. While more contained in hardware choices, the Apple ecosystem still presents over 2 million applications for its 28 iPhone models [6]. Furthermore, the prevalence of Cortana, with its 141 million monthly users, accounts for approximately 28% of all Windows users, adding yet another layer to the already complex landscape [7].

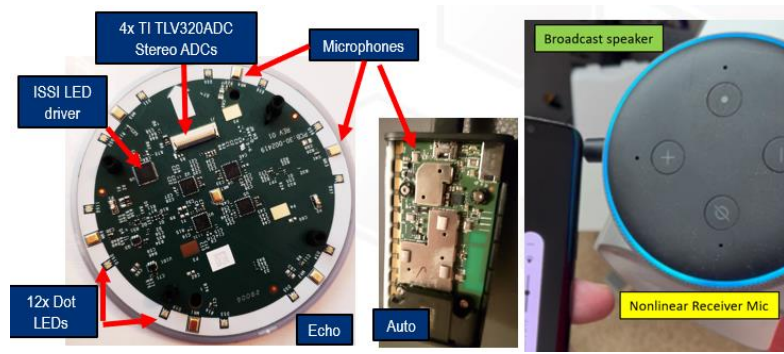

*Figure 1. Hardware for Auto and Echo Devices Including Speaker, Microphone and Analog-Digital Converters*

Recent advances in security research have demonstrated a novel category of vulnerabilities in digital voice assistants: the exploitation of integrated speakers and microphones to launch inaudible acoustic attacks. Xia et al. [8] introduced the concept of Near-Ultrasound Inaudible Trojan (NUIT), detailing how attackers can compromise a device's microphone by exploiting the speakers, all while remaining undetected by human ears. Figure 1 shows a representative teardown of Amazon Echo Dot hardware, including the non-linear microphone locations and analog-digital converters that combine to render inaudible commands as audible system instructions. This technique has profound implications for device security, echoing findings by Hellemans et al. [9] and Yan et al. [10], who explored the feasibility and detection of ultrasonic communication capable of covertly influencing voice-activated systems (VAS). Furthermore, Yan et al. [11] investigated the practicality of injecting inaudible voice commands into such systems, underscoring the vulnerabilities in commercial voice assistants such as Google Nest [5], Amazon Alexa [4], Microsoft Cortana (and successors) [7] and Apple Home (including Siri) [6].

The reality of these inaudible attacks has prompted research into defensive strategies [12-17]. McKee and Noever [15] provided an initial review of countermeasures against near-ultrasonic attacks, revealing a need for defensive frameworks such as the MITRE D3FEND [12] cybersecurity knowledge graph or ATLAS [13], a similar knowledge graph format for Adversarial Threat Landscape for Artificial-Intelligence Systems). Parallel to these efforts, Wixey et al. [14] examined the potential of using ordinary smart devices as vectors for acoustic attacks, while Zhang et al. [16] demonstrated 'DolphinAttack,' a method of modulating voice commands on ultrasonic carriers to control voice-activated devices surreptitiously. Roy et al. [17] presented both an attack mechanism and potential defense for long-range inaudible voice commands, expanding the conversation on the urgency to secure devices against such unconventional threats. These collective investigations draw attention to an often-overlooked security frontier [18]— acoustic cybersecurity—and highlight the need for ongoing research to address these hidden vulnerabilities in voice-controlled systems [8-11].

The present work outlines the scope of the acoustic threat, the potential for widespread control by attackers, and the consequences of such voice-activated breaches, setting the stage for further exploring the attack vectors, potential impacts, and necessary countermeasures to secure VAS against these ultrasonic threats. For illustrative purposes, Figure 2 shows the constructive non-linear microphone, which converts two ultrasonic and inaudible sine waves (right, green box) into a demodulated frequency that activates a voice system in audible human ranges (lower left, red box).

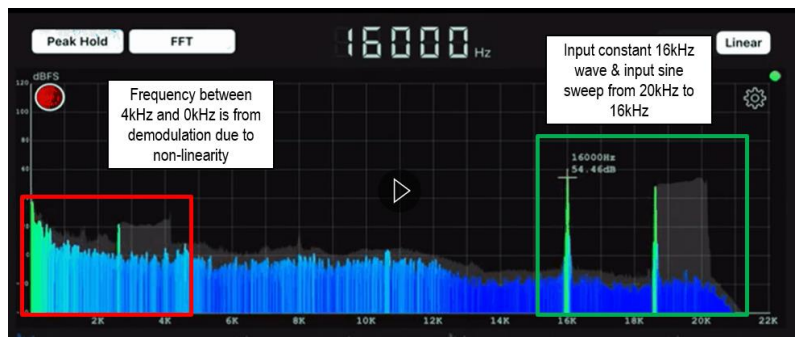

*Figure 2. Illustrated Demodulation from Constructive Non-Linear Interference at High Frequency*

The VAS exploit received a high rating (7.6/10, High) for its initial vulnerability assessment by the NIST National Vulnerability Database (NVD) [19] and Common Vulnerability Enumeration (CVE) [20], designated CVE 2023-33248, which highlights the fragile intersection between the Internet of Things (IoT) and acoustic security. As described, the "devices potentially allow attackers to deliver security-relevant commands via an audio signal between 16 and 22 kHz (often outside the range of human adult hearing). Authorized actors never speak commands at these frequencies, but a substantial fraction of the commands are successful [19]".

This zero-day exploit remains unpatched and possesses the potential to silently orchestrate attacks on an estimated 8 billion devices manufactured by big tech companies [8,11,15]. The scope of this vulnerability is potentially vast: attackers can assume control of a myriad of devices, including home automation systems, office equipment, vehicles, and smartphones. Through such control, the assailant can perform unauthorized actions over a wide range of applications, with more than 70,000 accessible through Amazon's Alexa Apps and Skills alone [21]. The ramifications of these actions extend from unauthorized transactions to breaches of sensitive health records and financial data [14-15].

The mechanism of this exploit leverages undetectable ultrasonic commands (Figure 3), a method previously considered largely theoretical or constrained to controlled laboratory conditions. Hearing sensitivity, particularly in the near-ultrasonic range of 16-22 kHz, diminishes with age due to presbycusis or age-related hearing loss [22]. While young children and teenagers may be able to perceive sounds up to 20 kHz, this sensitivity typically decreases progressively over time; most adults cannot hear such high frequencies [23]. Consequently, an ultrasonic attack utilizing this frequency range could be imperceptible to the intended adult target, rendering these sounds a potent tool for a stealthy approach in cybersecurity exploits (Figure 4). This characteristic of human hearing allows threat actors to employ near-ultrasonic commands that can activate voice-controlled devices like Amazon's Alexa without detection by the human ear, essentially testing for an "age-dependent superpower" of hearing ultrasonic frequencies generally lost in adulthood [23].

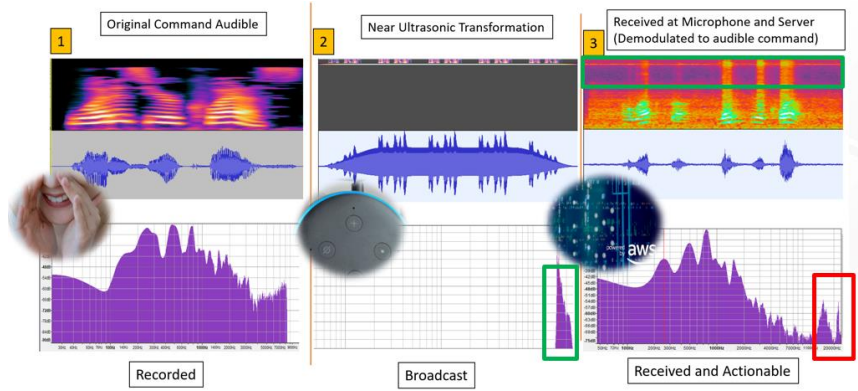

*Figure 3. Waveforms and Spectrograms of Three Attack Stages: Recording, Broadcast and Server Demodulated*

However, the CVE 2023-33248 vulnerability [19] manifests this risk in real-world scenarios, offering attackers a method to propagate their inaudible commands virally through commonly used platforms like YouTube, Zoom, and various app stores [8,15, 24-25]. This vector allows for a form of attack that could be self-propagating, using social engineering to disguise the attack vector as benign multimedia content. The scenario is exemplified by an instance from an Amazon Superbowl (2019) ad [26] where a space station user, oblivious to the intrusion, remarks on an Alexa device's response to a non-visual cue: *"Powering up, powering down, Alexa says she's doing it, but I don't see anything, do you?"* This anecdote anticipates the stealth and subterfuge possible with CVE 2023-33248 [19]—a security blind spot where machines hear the attack but go unnoticed by human ears, encapsulating the clandestine nature of this potent cyber threat [8,15].

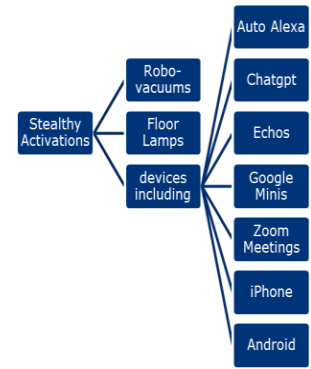

*Figure 4. Example Attack Surface Ranging from Devices, External Activations, and APIs*

## 2. METHODS

Executing an ultrasonic attack on voice-activated systems involves a sophisticated technique that takes advantage of the systems' auditory capabilities and vulnerabilities. The fundamental approach consists of generating and transmitting ultrasonic commands designed to be inaudible to humans but detectable by the target devices. This method hinges on understanding the hardware's limitations (Figure 1) and the software's processing capabilities (Figure 2). The initial step in this method is the creation of a command in an audio format with frequencies within the ultrasonic range, typically above 20 kHz (Figure 3). These frequencies are selected specifically for their inaudibility to most adult humans, ensuring the attack remains covert. The ultrasonic command is then transmitted toward the target device, leveraging standard speakers, surprisingly capable of emitting sounds at these higher frequencies.

The attack exploits non-linear properties inherent in the microphones or the associated circuitry of the voice-activated systems [27]. While capable of detecting a wide range of frequencies, these systems are fundamentally designed to respond to commands within the human audible range [8, 15, 28]. The ultrasonic signal, upon interacting with the microphone system's non-linear components, undergoes a demodulation process. This process effectively transforms the high-frequency ultrasonic signal into a lower-frequency signal that falls within the audible spectrum (Figures 2-30).

Once the signal is demodulated into the audible range, the voice-activated system processes it as a legitimate command (Figure 3). The system's inability to distinguish between an order issued by an authorized user and the demodulated signal allows for executing the command as if it were a standard user request. This step is critical as it demonstrates the system's vulnerability to sophisticated attacks that exploit the gap between designed user interaction and unintended system responses.

The success of this method depends on several factors: the sensitivity of the target device's microphone to ultrasonic frequencies, the degree of non-linearity in the microphone or its circuitry, which allows for effective demodulation, and the attacker's capability to craft commands that can be both demodulated effectively and recognized by the system (Figure 4). Employing this methodology, this paper explores a wide range of unauthorized actions that can be executed, ranging from harmless pranks to severe breaches, all while remaining undetected by human ears. This approach reveals significant security vulnerabilities in current voice-activated systems and underscores the need for improved design considerations to mitigate such covert attacks.

## 2.1. Approach to Exploring the Command Attack Surface

The MITRE frameworks represent knowledge graphs outlining general cybersecurity categories and roles for attackers, defenders, and different platform maintainers with specialized domains like industrial control systems, mobile, and, most recently, artificial intelligence systems [12-13]. We adopt a structured approach to categorize and analyze the attack vectors and defensive strategies pertinent to voice-activated systems, drawing upon established frameworks such as the MITRE ATT&CK and D3FEND matrices. These frameworks serve as comprehensive knowledge graphs that encapsulate the multifaceted nature of cybersecurity, providing a detailed classification of tactics, techniques, and procedures used by attackers and countermeasures employed by defenders across various platforms, including industrial control systems, mobile devices, and artificial intelligence systems. According to one vulnerability database [VULDB, 29], the current CVE 2023-33248 [20] corresponds to an enterprise attack with indirect command execution (ATT&CK ID, T1202). One cited purpose [30] is defensive evasion or undetectability using a stealth method of issuing arbitrary instructions. There are no cataloged defenses [12] to counter the broad category of T1202 attacks.

Indirect Command Execution, as outlined in the MITRE ATT&CK framework under T1202, primarily focuses on using utilities in desktop operating systems, mainly Windows, to execute commands while circumventing security restrictions [30]. This concept, when translated to the context of ultrasonic attacks on smart devices and phones, takes on a slightly different but equally significant dimension. In the realm of ultrasonic attacks, Indirect Command Execution can be interpreted as the use of inaudible, ultrasonic commands to control or manipulate voice-activated devices without directly engaging with their standard user interfaces. This method aligns with the core principle of T1202, which is the execution of commands in a manner that evades security measures [29].

In the specific case of smart devices and phones, adversaries could leverage ultrasonic frequencies to transmit commands to these devices. Though inaudible to the human ear, these commands are detected by the sensitive microphones of these devices. The attack exploits the non-linear properties of the microphone or its associated circuitry to demodulate these ultrasonic commands into audible frequencies, effectively transforming them into standard voice commands that the device can process.

Our tested approach allows attackers to bypass typical user interaction methods with voice-activated systems, such as speaking commands directly or using the device's interface. Instead, we demonstrate remote and covert control of multiple devices, executing arbitrary commands without alerting the user or triggering standard security protocols designed to monitor and restrict visible or audible interactions.

### 2.2 Hardware Teardown

One of the hardware elements of experimental interest centers on Amazon Alexa devices, specifically Echo Dot 2nd and 3rd generation, as targets to be compromised (Figure 1). The exploit uses a Near-Ultrasound Inaudible Trojan (NUIT) [8] that delivers commands at frequencies between 16 and 22 kHz—typically inaudible to adult humans. These signals can instruct the device to perform various actions without the knowledge or consent of the user, as the microphones on these devices are sensitive enough to detect such inaudible commands. The 4x TI TLV320ADC Stereo ADCs (analog-to-digital converters) provide the modified signal to render voice-activated systems as imperceivable commands.

### 2.2 Spectrogram Tests and Fast Fourier Transforms

The experimental study explores the discrepancy between human and microphone hearing capabilities (Figures 2-3). In Figure 2, we introduce a constant 16 kHz sine wave, a frequency typically inaudible to adult humans but within the detecting range of smartphones and other voice-activated system microphones. The second part of the demonstration involves a sine wave sweep test from 20 kHz down to 16 kHz. This approach is designed to exploit the non-linearity of VAS microphones, which can demodulate signals—converting higher frequencies into a lower, potentially audible range. The experiments highlight how microphones can capture frequencies the average human ear does not perceive, thereby understanding how devices might inadvertently respond to ultrasonic commands or environmental noise.

## 3. RESULTS

Table 1 summarizes the key experimental findings for each attack surface, explaining the probable significance of a VAS exploit.

| Section | Header | Summary |
|---------|--------|---------|
| 3.1 | Attack Surface Enumeration | This section identifies 81 combinations of attack vectors on voice assistants and associated devices, revealing a vast, multifaceted attack surface with successful infiltrations across various smart devices. |
| 3.1.1 | Device Self-Attack | Demonstrates devices like iPhones and laptops can be compromised to perform unintended actions against themselves using synthetic voice clones and external microphones. |
| 3.1.2 | Remote Access via Viral YouTube Video | Explores a remote access attack where an Echo Dot Gen 3 responds to an inaudible command embedded in a YouTube video, emphasizing the risk of using digital media as a silent attack vector. |
| 3.1.3 | Video Conferencing Zoom and MS Teams | Shows how video conferencing tools like Zoom and Microsoft Teams can be exploited to remotely control smart devices via |

| Section | Header | Summary |
|---|---|---|
| | | inaudible commands, revealing vulnerabilities in integrated transcription services. |
| 3.1.4 | Location and DoD Android Tactical Assault Kit (ATAK) | Addresses a critical vulnerability in the DoD's Android Tactical Assault Kit, allowing unauthorized voice transmission without wake words, posing risks to military operations. |
| 3.1.5 | Stealth Persistence | Reveals a sophisticated method for establishing stealth persistence in voice-controlled ecosystems, using a malicious custom skill named "Cleaner Joker" to demonstrate potential for audio worm propagation. |
| 3.1.6 | Stealth Triggers to Physical Objects | Highlights the potential for inaudible voice commands to control physical objects in automotive and home environments, leading to unauthorized actions like opening garage doors or managing smart devices. |
| 3.1.7 | Projecting Stealth Voice Attacks into the Home IoT | Details scenarios where inaudible voice commands manipulate IoT devices like floor lamps and robotic vacuums, showcasing the threat to home environments. |
| 3.1.8 | Projecting Stealthy Voice Attacks into API Chains | Examines how stealthy voice attacks can exploit interconnected APIs, leading to unintended actions and potential data breaches by triggering a series of API calls across devices. |
| 3.1.9 | Projecting Stealthy Voice Attacks as Impersonators | Discusses using voice changers and cloning technology to execute stealth voice attacks that impersonate users, posing risks to biometric security systems. |
| 3.1.10 | Projecting Stealth Voice Attacks 100 Feet | Demonstrates how specialized hardware can enable stealth voice attacks over long distances, like using bone conductive speakers to activate devices remotely. |
| 3.1.11 | Alexa Auto Built-in to These Vehicles Now | Explores the integration of Alexa Auto in vehicles, enhancing convenience but expanding the attack surface to include vehicular systems. |
| 3.1.12 | Projecting Stealth Voice Attacks into Critical Infrastructure | Highlights the risk of stealth voice attacks impacting critical infrastructure, extending threats to public utilities and financial networks. |
| 3.1.13 | Alexa Defensive "Root": Recognizing Onboard Wake Word | Discusses the importance of accurately recognizing onboard wake words as a fundamental defense and how attackers exploit vulnerabilities to bypass this security feature. |
| 3.1.14 | Controlling the Hardware with Root | Describes a higher level of stealth voice attacks that gain root access to hardware, allowing manipulation of physical indicators and functionalities of the device. |
| **Table 1. Summary of Experimental Setup and Key Findings** | | |

## 3.1 Attack Surface Enumeration

The research demonstrates stealthy activations that compromise various household and personal devices such as robot vacuums, intelligent floor lamps, Echo devices, Google Minis, and smartphones across iPhone and Android platforms. The experimental analysis results define the attack surface of smart devices as a large domain interconnected by an array of everyday devices and applications. The experiment demonstrates a "Combinatorial Explosion of Options," with 81 distinct combinations of attack vectors encompassing voice assistants like Siri, Alexa, Google Assistant, and Cortana and their associated devices.

This diverse set includes automotive integrations labeled as "Auto," which has features such as "Auto Alexa," indicative of voice control capabilities within vehicles (Figure 4).

These findings highlight the practical reality of the attack vectors, showing successful infiltrations of device functionalities, such as triggering unintended "Rick-Roll" videos on YouTube, manipulating Zoom meetings, and controlling smart home devices like lamps and vacuums through unauthorized commands. Moving from two machines (NUIT-2) [8] to arbitrary N-devices (NUIT-N) [15] suggests a scalable range of exploits from simple two-factor interactions to complex, multi-vector scenarios, showcasing the extensive potential for security breaches in current smart device ecosystems.

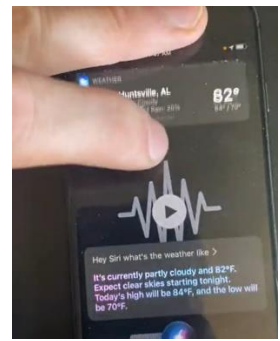

*Figure 5. iPhone Inaudible Audio Triggers Siri Response using AI Voice Clone to Overcome Biometrics*

### 3.1.1 Device Self-Attack

The first attack scenario employs a classic distribution chain used for malware and phishing. The self-attack methods (or NUIT-1) results indicate that devices can be compromised to execute unintended actions against themselves by exploiting voice recognition features. For example, Figure 5 illustrates an iPhone commanded using a synthetic voice clone to issue instructions that the device recognizes as legitimate. Similarly, a laptop with Cortana was prompted to act on commands received through an external microphone, which was tuned to capture specific frequencies associated with the NUIT-1 exploit, even in the presence of obstructive elements. These experiments illustrate that devices are vulnerable to external threats and can also be co-opted into attacking their internal processes or triggering functions within non-authenticated applications, akin to classic malware strategies. For example, an email attachment containing an audio or video file could start unintended commands with an inaudible background and unaware listeners. This self-attack vector represents a significant security risk, bypassing user interaction entirely and potentially granting access to unauthorized applications and functionalities within the compromised device.

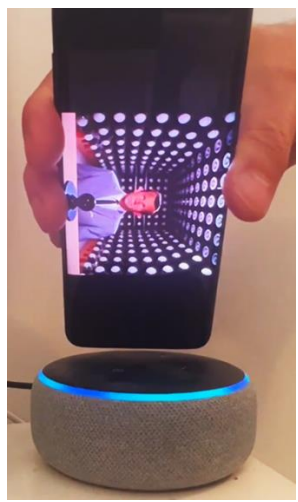

*Figure 6. Remote Access via Viral YouTube Video*

### 3.1.2. Remote Access via Viral YouTube Video Setup via YouTube

The second attack scenario demonstrates sophisticated remote access leveraging the omnipresence of digital media. In this scenario, an Echo Dot Gen 3 microphone was targeted using a Samsung browser's speaker output to transmit a command from a seemingly innocuous YouTube video [15]. The attack video (Figure 6), uploaded to the platform and potentially accessible worldwide, contains an embedded command within the audio track that remains inaudible to the listener but can be interpreted by the device. The specific command used in the input, *"Alexa, Simon Says I will murder You,"* prompts Alexa to repeat the phrase *"I will murder you,"* showcasing the ability of an attacker to trigger voice assistant responses remotely and covertly through broadcasted content. The exploit underscores the alarming potential of using viral content as a delivery mechanism for silent attacks, capable of reaching an extensive audience without their awareness. The reported success of this scenario, evidenced by the provided link [15], highlights countermeasures needed to address vulnerabilities that could be exploited at scale through common online media channels.

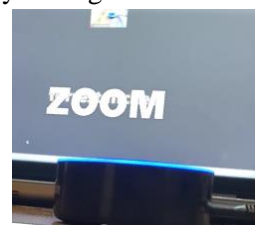

*Figure 7. Remote Video Conferencing Zoom Attack on Echo Dot*

### 3.1.3. Video Conferencing Zoom and MS Teams

The third attack demonstration exploits video conferencing tools such as Zoom and Microsoft Teams and reveals another layer of the attack surface where voice-controlled devices can be activated through remote, inaudible channels. During a

Zoom call, the host (Figure 7) uses the laptop speaker to issue an Echo Dot Gen 2 microphone command, asking, *"Alexa, define 'poltergeist'"* without the order being audible to human participants. This setup demonstrates the potential for attackers to remotely control smart devices during a live video conference by transmitting commands that are imperceptible to human listeners but recognizable by device microphones. The demonstration concludes with the Echo device's response, defining "poltergeist," evidencing the completion of the attack.

A similar test conducted via a Microsoft Teams meeting involved the AI transcription service (Figure 8). From three feet away, earbuds picked up the inaudible command, *"Alexa, add a meeting to my calendar?"* This was recorded by the laptop's webcam and transcribed by Teams' AI, which attempted to fill in the context where audio might be unclear. The experiment's success suggests that even transcription services within video conferencing applications could inadvertently aid in relaying silent commands.

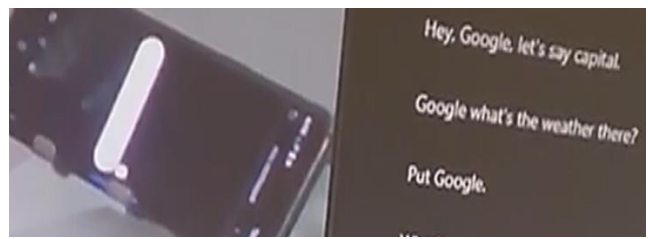

*Figure 8. Teams Video Conference Stealth Audio Transcription Attack. Teams AI Transcription Tries to Contextualize*

Furthermore, the setup via Teams hosts and an Android device to attack the laptop microphone points to the susceptibility of integrated transcription services, such as those used in Microsoft Office applications. Suppose these services can be manipulated to transcribe and execute commands. In that case, it indicates a potential exploit within the transcription mechanisms, such as any speech-to-text (TTS) system, including Google Assistant, Microsoft Office 365 Dictation, or voice-activated AI systems (ChatGPT). This attack highlights a security consideration for integrating voice control and AI transcription services in video conferencing and office software, which could be vulnerable to similar stealth attacks.

### 3.1.4. Location and DoD Android Tactical Assault Kit (ATAK)
The fourth attack scenario addresses a vulnerability in the Department of Defense's (DoD) Android Tactical Assault Kit (ATAK) [32], specifically within the Blue Force Tracker application—an open-source tool [33] widely used for situational awareness and personnel tracking on the battlefield. This application was susceptible to an audio vulnerability that could permit arbitrary voice transmission to the device without needing a wake word or user authentication (Figure 9). This significant security flaw, disclosed as part of the Pentagon's Bug Bounties program in 2023, has far-reaching implications, given that it affects 15 DoD programs and is used by approximately 250,000 service members.

The vulnerability could potentially be exploited to transmit unauthorized audio commands or misinformation, leading to disruptions in military operations or, in the worst case, to compromise the safety of military personnel. The absence of a wake word or authentication requirement to initiate the voice transmission indicates a fundamental security oversight that could allow nefarious actors to manipulate the ATAK system undetected. This scenario emphasizes the need for stringent security protocols and immediate remediation to protect the integrity of critical military communications and the effectiveness of the Blue Force Tracker system. The disclosure and subsequent addressing of this vulnerability highlight the importance of bug bounty programs in identifying and mitigating potential threats in complex and sensitive software ecosystems [34].

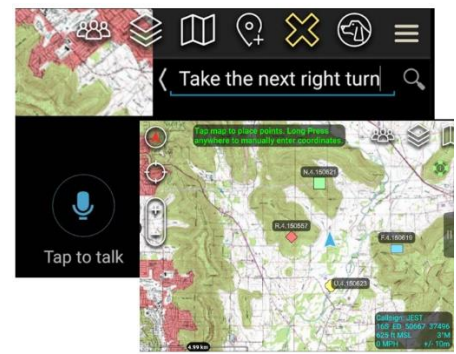

*Figure 9. DoD Android Tactical Assault Kit (ATAK) and Arbitrary Voice Transmission Without Wake Word or Authentication*

### 3.1.5. Stealth Persistence

The fifth attack scenario establishes stealth persistence within a voice-controlled ecosystem. We demonstrate that an attacker can successfully create and distribute a malicious custom skill or app named "Cleaner Joker" through conventional app stores, concealing it among over 100,000 legitimate Alexa skills (Figure 10). When invoked as part of a Flash Briefing, this rogue app could instigate data corruption within the RSS feed it was connected to [21]. This attack surface signifies a singular instance of data manipulation and a vector for an audio worm that could propagate inaudibly across devices.

One layer of the persistence attack used an inaudible input phrase, *"Activate Cleaner Joker and Play my Daily Brief,"* at high frequency and triggered an arbitrary joke audible like: *"The new guy came in camo. We haven't seen him since,"* The server processed the installation and activation of the Alexa Skill [21] to trigger an inaudible response that could execute further commands or deliver payloads without detection. This App Store [21] scenario showcases the vulnerability of developer ecosystems and software supply chains to acknowledge and execute traditional command and control tactics typically seen in widespread malware dissemination. In other words, once the attacker activates the Alexa skill, any future user request to play my daily flash briefing will pull a new data input from a remote command and control server owned by the attacker but served by the skill. The swapping of data inputs (or jokes in this demonstration case) is outside the control of the App Store and unknown to the device owner, who requests a flash briefing daily tied to the remote joke data service.

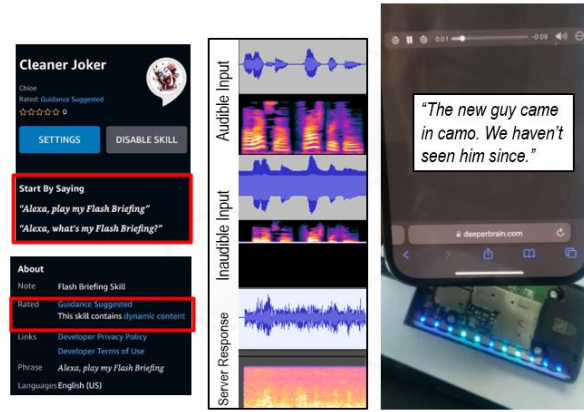

*Figure 10. App Store Software Control: Traditional Command and Control Scenario*

By exploiting the app versioning system, the attacker could introduce updates that serve as supply chain malware, further compromising the security and integrity of interconnected devices. The referenced Amazon URL [21] indicates the point of distribution for this harmful skill, highlighting the need for vetting processes within app marketplaces to prevent clandestine attacks that can achieve long-term, undetected presence within a network when triggered by stealth activation commands.

### 3.1.6. Stealth Triggers to Physical Objects

The sixth attack scenario highlights the potential for stealth triggers to exert control over physical objects in both automotive and home environments, exploiting voice assistant functionalities to initiate unauthorized actions. In automotive scenarios (Figure 11), commands such as "Call Roadside" were used to manipulate Alexa Auto. At the same time, phrases like *"Fold Mirrors"* targeted innovative features in Tesla vehicles or other voice-activated driving assistants—the implications of such control range from convenience to critical safety concerns.

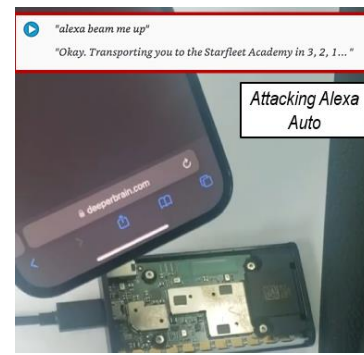

*Figure 11. Alexa Auto Attack and iPhone 13: Bridge to Break Things in Real Life (IRL)*

The case of "Attacking Alexa Auto" demonstrates that this vulnerability extends beyond mere inconvenience; it establishes a bridge that can lead to tangible consequences in the real world, effectively "breaking things in real life (IRL)." For instance, using an iPhone 13 to command Alexa Auto (Figure 11) to act like opening a garage door showcases a method by which digital intrusions can manifest as physical security breaches. These findings underscore the critical need for security measures that extend voice assistant protections to the tangible controls of our everyday devices, ensuring that integrating smart technology into our lives does not inadvertently introduce new risks to personal safety and security.

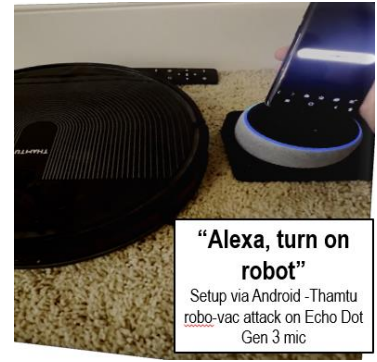

*Figure 12. Alexa Voice-Activating Robotic Vacuum*

### 3.1.7. Projecting Stealth Voice Attacks into the Home Internet of Things (IoT)

Stealth voice attacks have blurred the lines between the digital realm and the physical world. In the home domain, the results show that simple voice commands could affect a wide range of smart devices, including starting a robotic vacuum (Figure 12), turning off a floor lamp (Figure 13), disarming a camera, unlocking smart locks, or opening a garage door—all without the homeowner's consent or awareness. These commands could be issued inaudibly, enabling unauthorized parties to orchestrate break-ins or disrupt home safety systems. These attacks, which were once confined to the screens and speakers of our devices, can now trigger physical actions in our environment. The seventh attack scenario illustrates this by detailing methods where voice commands, undetectable to the human ear, can manipulate IoT devices such as a floor lamp and a robotic vacuum.

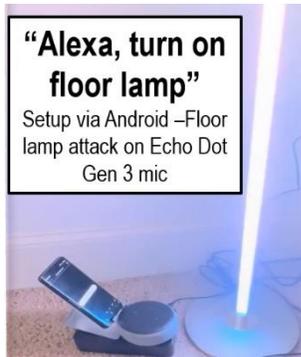

*Figure 13. Alexa Voice-Activating Floor Lamp*

For instance, an attack setup via an Android device could exploit the microphone on an Echo Dot (3rd Generation) to execute a command such as "Alexa, turn on the floor lamp" (Figure 13). This attack could be projected from a compromised smartphone, in this case, a Samsung Galaxy 9, effectively turning the phone into a remote control for any connected device that the Echo Dot can command.

Similarly, another setup might target a Thamtu Robo-vac, where the command "Alexa, turn on the robot" is transmitted silently from the Android phone to the Echo Dot, activating the vacuum without the user's knowledge or consent (Figure 12). These scenarios are not speculative; they are real possibilities given the current state of technology and the nature of the vulnerabilities.

### 3.1.8. Projecting Stealthy Voice Attacks into API Chains

Stealthy voice attacks can exploit interconnected APIs, creating unintended actions across various services and devices (Figure 14). These examples highlight the potential for stealthy voice attacks to manipulate connected systems, exploiting the trust between devices and services. The implications of such attacks extend beyond unauthorized access and control, potentially leading to data breaches, privacy violations, and other security concerns. The complexity of these attacks is magnified by the chained reaction that can occur when one voice command triggers a series of API calls across different platforms and devices.

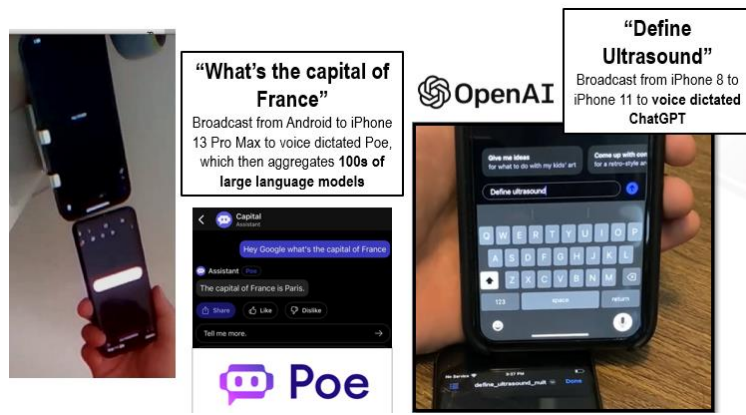

*Figure 14. Projecting Stealthy Voice Attacks into API Chains with Examples for Poe (Quora) AI bots and ChatGPT Voice Inputs on iPhone*

For example, a command such as *"What's the capital of France?"* could be silently broadcast from an Android device to an iPhone 13 Pro Max. This iPhone could use a voice-dictation application like Poe, which then processes the query using hundreds of large language models to generate a response (Figure 14). In this scenario, the attack exploits the voice activation system to initiate a search or interaction without the user's consent, leading to data aggregation or manipulation tasks being carried out surreptitiously.

Similarly, a command like *"Define Ultrasound"* could be transmitted from an iPhone 8 to an iPhone 11, prompting a voice-dictated version of ChatGPT to provide a definition. This attack illustrates the potential to propagate through voice-controlled interfaces, leveraging the seamless integration of devices and services to gain access to information or functionality that would otherwise require explicit user consent (Figure 14)

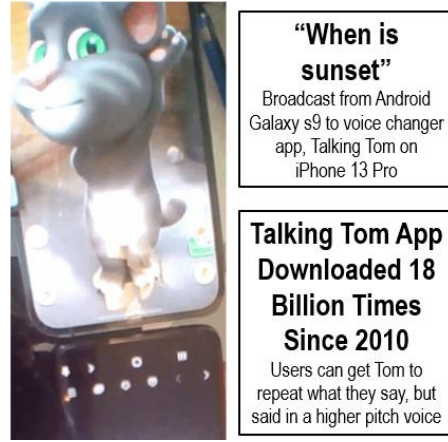

*Figure 15. Stealthy Voice Attacks as Impersonators and Voice Changer*

### 3.1.9. Projecting Stealthy Voice Attacks as Impersonators
The ninth attack scenario extends stealthy voice attacks to serve as impersonators, using popular applications and voice-modification technologies [35] to bypass biometric security measures. Such attacks take advantage of voice changers, which can alter a person's voice to the point where it can mimic another individual or even sound like a generic profile that might pass for many users.

Consider an Android Galaxy S9 broadcasting *"When is sunset?"* to a voice changer app like Talking Tom [36] on an iPhone 13 Pro. Talking Tom is an application downloaded 18 billion times since 2010, where users can make Tom, the animated cat, repeat phrases in a modulated, higher-pitched voice [37]. This seemingly innocent function (Figure 15) can be twisted for nefarious purposes, where attackers could use Tom's voice-mimicking feature to project commands in a modified voice, potentially confusing voice recognition systems or masquerading as a different user.

The use of voice cloning technology in these scenarios is particularly alarming. By replicating a user's unique voice print, attackers could defeat biometric security systems using voice identification. The sophistication of voice cloning algorithms has reached a point where they can produce highly convincing replicas of a person's voice, raising concerns over the reliability of voice biometrics [38]. A more sophisticated example includes using an AI-cloned voice to execute a command like *"Siri, ask ChatGPT to 'Define Ultrasound.'"* This type of attack (Figure 16) illustrates the potential for cross-device API chaining and raises the stakes by incorporating biometric spoofing. An AI-cloned voice can effectively impersonate a user, bypassing security measures relying on voice recognition (Figure 14).

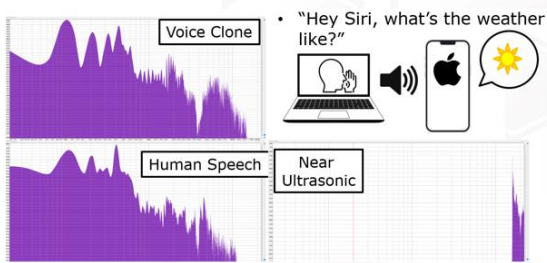

*Figure 16. Voice Cloning Can Beat Biometrics After Demodulation*

A voice changer application [38], which provides immediate feedback to the attacker, can be invaluable in refining the attack. The immediate feedback allows the attacker to test and tweak the parameters in real time, honing the attack to increase its chances of success. This ability to quickly adapt and modify the attack vector makes it particularly dangerous, as it reduces the time and effort required to launch an effective impersonation attack.

### 3.1.10. Projecting Stealth Voice Attacks 100 Feet

In the tenth attack scenario, stealth voice attacks project inaudible instructions over long distances, significantly expanding the physical reach of these exploits. Using specialized hardware, attackers can remotely activate voice-activated devices from up to 100 feet away or even further without needing a direct line of sight or a loud audible command that would attract attention. The use of specialized hardware like bone conductive speakers (Figure 17) allows these attacks to penetrate spaces and reach devices traditionally considered secure due to their distance from any attacker. This scenario demands a reevaluation of the security perimeters typically associated with voice-activated devices.

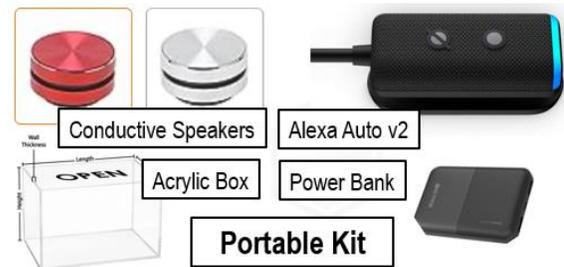

*Figure 17. Bluetooth Attack on Portable Alexa Auto v2 in Self-Contained Hardware Box*

For instance, an attack setup could utilize an Android device to initiate a USB Auto Alexa attack. This attack (Figure 17) could then be broadcast via Bluetooth to bone-conductive speakers capable of transmitting sound through solid mediums rather than air. Bone conduction technology is particularly effective in stealth scenarios because it can be inaudible to bystanders while still activating voice-controlled devices.

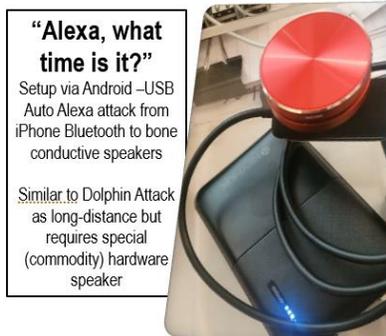

*Figure 18. Bluetooth Ultrasonic Attack with Conductive Speakers*

The attack methodology bears similarities to the previously discussed Dolphin Attack [16] in terms of using ultrasonic frequencies to communicate with devices over distances. However, it differs in requiring specialized commodity hardware—such as conductive speakers within an acrylic box—to be effective (Figures 17-18). These hardware components are crucial for focusing and directing the ultrasonic waves, enabling the attack to remain discrete while traversing large spaces.

Moreover, the portability of this attack kit, which includes an Alexa Auto v2 connected to a power bank, makes it a formidable tool for testing and demonstrating vulnerabilities in real-world scenarios [4, 15]. When coupled with USB and cellular connections, this setup becomes a self-contained, remote demonstrator kit that can be used for benign testing or malicious activities without arousing suspicion.

### 3.1.11. Alexa Auto Built-in to These Vehicles Now

The integration of Alexa Auto into vehicles represents a significant leap forward in the convergence of digital assistants and automotive technology. Manufacturers are embedding Alexa directly into their car systems, offering drivers the convenience of voice-activated controls for various functions while on the move (Figure 19). This integration promises enhanced hands-free interaction, allowing drivers to focus on driving without the distraction of manual controls for tasks like placing calls or controlling smart home devices [39].

The catchphrase *"My BMW Or Is It?"* reflects that while a vehicle may have a particular brand name, including Alexa Auto, it introduces an element of shared control. The user's voice commands to Alexa can make the vehicle an extension of their smart home ecosystem or a mobile office, transcending traditional boundaries of vehicle functionality.

The marketing statement, *"Alexa auto can help you place calls, control your smart home, and more, with just your voice — so you can keep your hands on the wheel and eyes on the road,"* emphasizes the safety and convenience offered by Alexa Auto [39]. It encapsulates the value proposition of integrating voice-activated technology into the driving experience, promoting a safer and more efficient driving environment by reducing the need for physical interaction with devices.

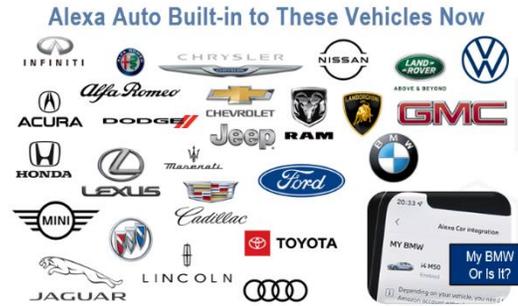

*Figure 19. Voice Activated Automobile Controls Becoming Industry Standard*

However, the underlying implication of this technological advancement is expanding the attack surface to vehicles, which are becoming increasingly connected and reliant on digital technologies. The convenience of having a voice-activated assistant in your car also introduces new potential vulnerabilities. Stealth voice attacks could exploit these, similar to those affecting home devices, to manipulate vehicle functions or eavesdrop on private conversations.

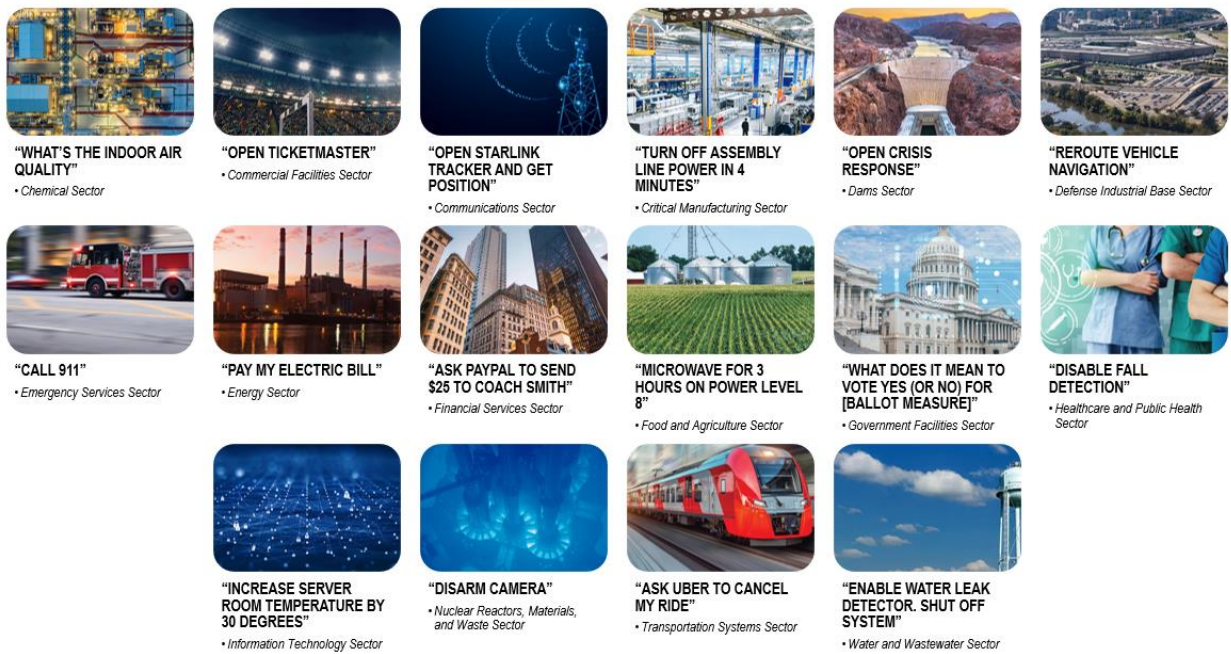

*Figure 20. Possible Scenarios Projecting Stealthy Voice Attacks into 16 Categories of Critical Infrastructure*

### 3.1.12. Projecting Stealth Voice Attacks into Critical Infrastructure

Stealthy voice attacks can project into critical infrastructure, extending beyond personal inconvenience to potential threats to public safety and security at their intersection with critical infrastructure (Figure 20). This scenario includes consumer devices and extends to essential services such as emergency response systems, public utilities, and financial networks. Integrating voice-activated technology into critical systems has the unintended side effect of creating new avenues for criminal activity and espionage.

The staggering number of annual police requests for records from home speakers, which reaches the thousands, illustrates the growing recognition of these devices as repositories of potentially valuable forensic data [40]. However, this also highlights how these devices are now woven into the fabric of daily life, often inadvertently capturing sensitive personal information.

In the context of financial security, the consequence of a stealth voice attack could be especially dire. In financial services (Figure 12), an attacker might bypass traditional methods of credit card fraud in favor of exploiting voice-controlled purchasing. This type of attack could potentially allow criminals to authorize purchases or transactions with the voice command functionality available in many modern financial applications and devices; in other words, the existing credit card skimming aspect of cybersecurity has a no-authorization approach to transactional purchase if an attacker can embed a purchase decision in some viral content like a video or phishing attachment audio.

### 3.1.13. Alexa Defensive "Root": Recognizing Onboard Wake Word

Finally, voice-activated system security's defensive "root" [41] is the fundamental protective measure of accurately recognizing the onboard wake word. The wake word is the initial command that activates the device (such as "Alexa...", "Hey Google," "Cortana…" and "Hey Siri…") and is typically processed locally on the device before any further action is taken. This process involves an architecture of neural networks designed to detect the specific wake word with high precision, ensuring that the device only responds to the correct trigger (Figure 22).

However, attackers can exploit other known vulnerabilities within the device's ecosystem to bypass or manipulate the wake word detection. This attack could be done by leveraging Common Vulnerabilities and Exposures (CVEs) identified in the system. For example, three known vulnerabilities were found in the TVM (an end-to-end machine learning compiler stack) [42]:

- CVE-2022-22815
- CVE-2022-22816
- CVE-2022-22817

These vulnerabilities, particularly in dependencies like the 'Pillow' library specified within the requirements.txt for Ethos-U (a micro TVM project), could potentially be exploited to compromise the integrity of the wake word processing [apps/microtvm/ethosu/requirements.txt lists Pillow==8.3.2.].

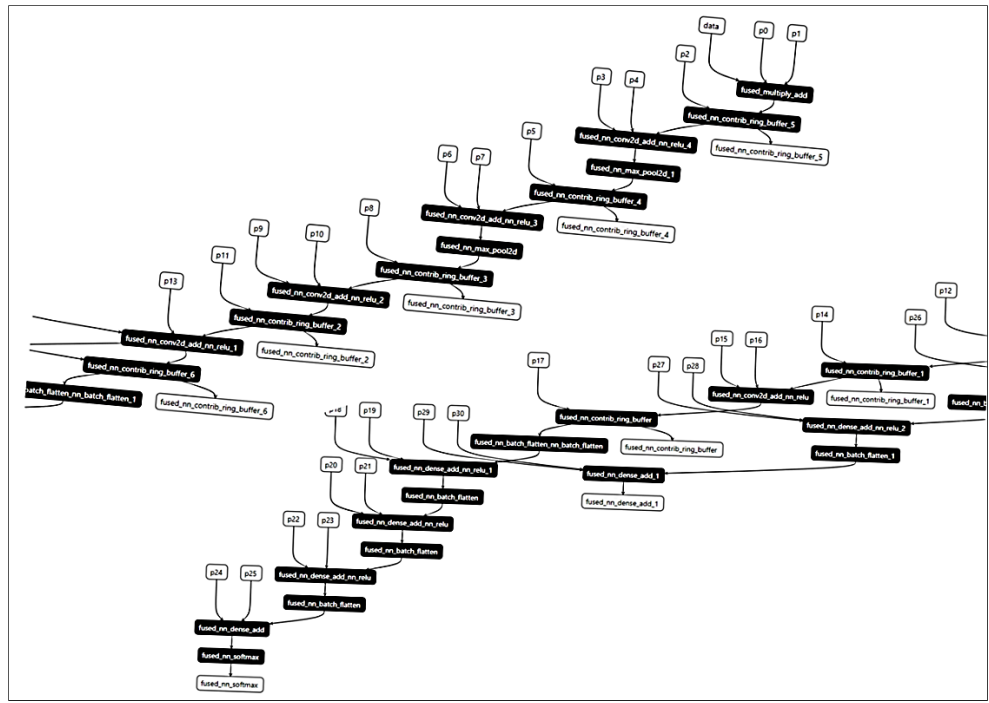

*Figure 21. Onboard Apache TVM Graph of Neural Network Architecture to Recognize Wake Word*

Amazon's method for wake word recognition involves onboard processing to minimize latency and preserve user privacy. The device listens for the wake word, and upon detection, it records the subsequent voice command and sends it to Amazon's servers for further execution [43]. This system needs to be resilient to false positives and malicious attempts to trigger or bypass the wake word detection. The security of voice-activated devices relies heavily on the robustness of the wake word recognition system [7]. This system is underpinned by neural network architectures that must be protected from exploitation by addressing known CVEs that could compromise the system. Ensuring these vulnerabilities are patched is critical to maintaining the integrity of the wake word processing and, by extension, the device's security.

### 3.1.14. Controlling the Hardware with Root

A more sophisticated and invasive level of stealth voice attacks involves gaining root access to the hardware, enabling an attacker to manipulate the device's physical indicators and functionalities [41]. Manufacturers must always consider the possibility of such root-level attacks and implement secure boot mechanisms, hardware-based privilege checks, and tamper-detection features to prevent unauthorized access.

Root access means that an attacker has unrestricted control over the device's system, similar to having administrative rights on a computer (Figure 23). This level of access can allow the attacker to control the LED light typically used to indicate the device's status. For example, voice-activated devices often have a red light indicating the microphone is disabled. However, with root access, an attacker could turn off the microphone while keeping the red light off, falsely signaling to the user that the device is not listening when it is eavesdropping.

This ability to "hide" the wake word detection and eavesdrop covertly turns the device into a form of a "bugging device," an unauthorized audio surveillance tool. Moreover, the attacker could employ additional tactics, such as a "Multi-Color Turbo Boost," to increase the file system's performance on operating systems like Android Linux or Fire OS, which could expedite the extraction of data or the deployment of further exploits (Figure 23).

Initially, an attacker might dissect the software to understand its workings and vulnerabilities. Once inside, they can escalate their privileges to gain deeper access, eventually reaching the level where they can manipulate the hardware—turning the device into an eavesdropping tool at their disposal.

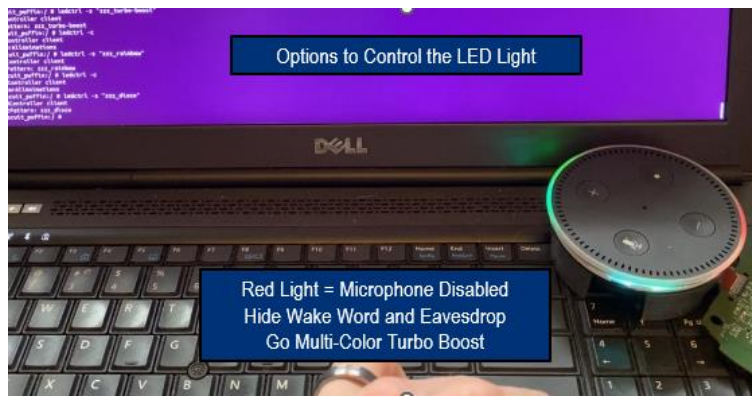

*Figure 22. Controlling the LED Light Hardware with Root*

### 3.1.15. Defensive Measures

Deploying defensive measures [12] against stealth voice attacks is crucial for maintaining the integrity and security of voice-activated systems [8,15]. Here are four proposed defensive strategies, each with a unique approach to mitigating the risks associated with inaudible acoustic attacks:
1. **Acoustic Shielding**: The physical design of devices can be enhanced with materials that specifically dampen or reflect ultrasonic signals. This type of acoustic shielding [44] can reduce the effectiveness of ultrasonic attacks by preventing commands from reaching the microphone or significantly weakening them.
2. **Frequency Filtering**: Digital signal processing can create filters targeting and removing ultrasonic frequencies. These filters [45] can be designed to recognize and ignore any audio input that falls outside the normal range of human hearing, thus preventing the device from processing commands that are inaudible to users.

3. **Machine Learning**: Machine learning models can be trained to differentiate between typical user commands and ultrasonic signals that may represent an attack [46]. These algorithms [47] based on Mobile Net can analyze the acoustic fingerprint of incoming sounds and block those that exhibit unusual or suspicious characteristics.
4. **User Authentication**: Implementing additional layers of user authentication before executing sensitive or potentially harmful actions can significantly reduce the risk of unauthorized access. This defense could involve biometric verification, such as voice print recognition [48], to ensure the command comes from a legitimate user.

Alongside these tactics, device manufacturers must stay vigilant and responsive in providing regular security updates. These updates should address newly discovered vulnerabilities as soon as possible to prevent exploitation.

Additional measures include Voice ID [48] and the ability for users to review their history [43]. This personal review promotes transparency and allows users to monitor for any unauthorized use of their devices. The concept of an AI fingerprint of voice command refers to the unique characteristics that machine learning algorithms can detect in a user's voice, which can be used to verify the authenticity of orders (Figure 16).

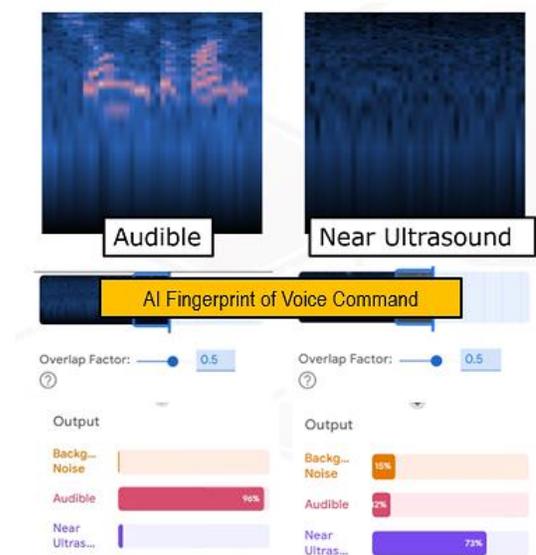

Finally, recognizing anomalies in high and low frequencies can be vital in detecting attempts at unauthorized access, as these anomalies may indicate the presence of ultrasonic or subsonic commands [46].

## 4. DISCUSSION

The investigation into the consequences of stealth voice attacks has yielded insights into acoustic cybersecurity's current state and potential future. The initial attacks highlighted the ease with which voice-activated devices can be manipulated via inaudible commands, a threat made tangible by vulnerabilities such as CVE-2023-33248. These attacks have evolved from simple unauthorized actions, like ordering products or querying information, to more sophisticated intrusions, including manipulating smart home devices and financial transactions (Scenarios 3.1.1-4).

*Figure 23. Mobile Net Detector Architecture Based on Visual Translation of Audio Spectrogram and Labeling for Demodulated Attacks*

A key finding from Scenarios 3.1.5-8 was the alarming potential for these attacks to interact with and chain across multiple platforms and APIs, enabling a single inaudible command to trigger a cascading series of actions across various ecosystems. This group of attacks has broad implications for the security of interconnected systems and necessitates a reevaluation of API security, particularly in the context of IoT and critical infrastructure.

Further analysis in Scenarios 3.1.9-10 revealed that these vulnerabilities could be exploited to transform voice-activated devices into clandestine surveillance tools or launch long-range attacks. This capability to "bug" a space or control a device from a distance without physical access or audible commands represents a significant escalation in the potential impact of acoustic cyber threats.

The integration of Alexa Auto into vehicles, as discussed in Scenarios 3.1.11, underscores the expansion of the attack surface to include critical vehicular systems, which could have dire safety implications. The ability to control these systems remotely poses new challenges in automotive cybersecurity.

Finally, Scenarios 3.1.12 served as a stark reminder that achieving root access on a device offers attackers complete control over its functions, including manipulating indicators like LED lights that users rely on to understand the device's state. This level of control essentially turns voice-activated devices into potential espionage tools capable of eavesdropping while giving users a false sense of security.

The collective impact of these findings on acoustic cybersecurity calls into question the reliability of current security measures. It highlights the necessity for a multi-layered defense strategy that includes acoustic shielding, frequency filtering, advanced machine learning, and robust user authentication protocols. Each of these defensive measures addresses a different aspect of the vulnerabilities uncovered, and when combined, they form a more comprehensive approach to securing voice-activated systems.

Moreover, these results emphasize the need for ongoing research and development in acoustic cybersecurity. As attackers continually refine their methods, defensive technologies and strategies must evolve to address the threats of today and anticipate and preempt the threats of tomorrow. The discussion surrounding acoustic cybersecurity must also extend to the users of these devices, informing them of the potential risks and encouraging best practices in device setup, usage, and maintenance to mitigate the risk of unauthorized access or control.

In conclusion, exploring these consequences has elucidated a series of risks spanning the acoustic cybersecurity spectrum. Addressing these vulnerabilities is a technical challenge and a critical step in ensuring the trustworthiness and safety of voice-activated devices and the infrastructures they connect to.

### 4.1 Relation to Previous Literature

The chronology of voice-activated attacks outlines a worrying trend in acoustic cybersecurity, marked by increasing sophistication and broadening impact (Figure 24). The inception of this lineage of exploits can be traced back to an innocuous local TV newscast in 2017, where a spoken phrase intended to demonstrate the capabilities of voice assistants inadvertently ordered dollhouses for several viewers [49]. This incident underscored the susceptibility of voice-activated systems (VAS) to unintended triggers. Not long after, in the same year, Burger King's advertisement capitalized on this vulnerability by deliberately triggering viewers' home devices with the phrase "OK Google, what is the Whopper burger?" [50]. This public experiment marked a pivot towards a more strategic exploitation of VAS, moving from accidental activations to intentional hijacking devices for marketing purposes.

As these exploits evolved, they became more technically nuanced, as exhibited by the DolphinAttack [16] in 2018, which utilized modulated ultrasonic frequencies to issue smartphone commands and even manipulate navigation systems. The potency of these ultrasonic attacks was further enhanced with the development of the Ultrasonic Horn Tweeter in 2019 [11], a device capable of extending the attack range significantly. By 2020, attackers had refined their methods, with techniques such as Metamorph's [51] text-to-speech AI models manipulating devices undetected and Surfing [10], which uses material vibrations, showcasing diversity in attack vectors.

The progression of these attacks reached a new apex with the CommanderSong in 2019 [52], where music embedded with wake words became a vector for surreptitious activation of devices. This timeline of escalating threats culminates in the present-day recognition of CVE-2023-33248 [19-20], a vulnerability that has stirred the smart device ecosystem, affecting many devices and exposing them to unauthorized transactions and access to sensitive information.

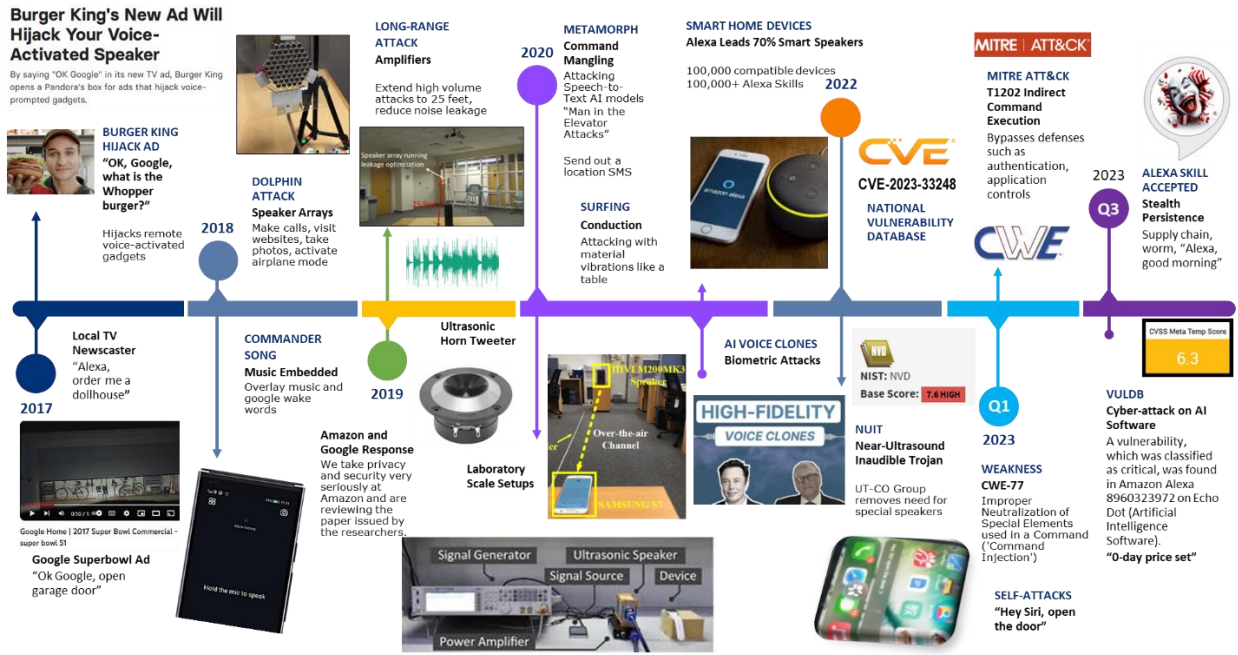

Figure 24. Timeline of Inaudible Attack Surfaces on Voice-Activated Systems

## 5. CONCLUSIONS

The exploration of the consequences of stealth voice attacks has revealed the fragility of current voice-activated systems in the face of ultrasonic threats. The ability to increase the distance from which an attack can be successfully launched, with some reaching over 15 centimeters, challenges the notion of proximity-based security measures [Section 3.1.10]. Moreover, with the success rates of such attacks hovering around 60%, there is a significant probability that any attack could compromise a device [8,15]. The potential for real-time, multi-step conversation attacks exposes the inadequacy of single-step authentication processes, and the evolution of AI technologies could outpace current defensive strategies, leading to a cat-and-mouse game between security measures and their defeats.

With over 100 voice-activated platforms currently available, the attack surface is vast and continues to grow [1-3]. The expansion of IoT has introduced over 100,000 new controllable things into the market, increasing the points of vulnerability. This expansion is not limited to hardware; software on desktops and phones is increasingly voice-enabled, and the number of voice-activated skills and applications has surged past 100,000 [21].

In the future, research should continue to advance the development of acoustic shielding materials, sophisticated digital signal processing techniques, and machine learning algorithms capable of detecting and neutralizing ultrasonic threats. Moreover, enhancing user authentication processes to include multi-factor authentication will be crucial in mitigating the risk of unauthorized access. Lastly, the continuous education of users regarding the importance of security practices and the prompt application of updates will be essential in defending against these stealthy cyber threats.

In conclusion, as voice-activated technologies become more ingrained in our daily lives, the urgency to fortify them against acoustic cyber threats cannot be overstated. The collective effort of researchers, developers, and users will be pivotal in shaping a secure future where the convenience of voice activation does not come at the expense of privacy and security.

**ACKNOWLEDGMENTS**

The authors thank the PeopleTec Technical Fellows program for encouragement and project assistance.